\documentclass[journal]{IEEEtran}
\pdfoutput=1

\usepackage{amsmath}
\usepackage{cite}
\usepackage[english]{babel}
\usepackage[normalem]{ulem}
\usepackage{standalone}
\usepackage{tikz}
\usetikzlibrary{shapes,shadows,arrows,positioning,calc}
\usepackage{multicol}
\usepackage{graphicx}
\usepackage{amssymb}
\usepackage{color}
\usepackage{bm}
\usepackage{soul}

\usepackage{nomencl}

\usepackage{ifthen}
\renewcommand{\nomgroup}[1]{%
	\ifthenelse{\equal{#1}{I}}{\item[\textbf{Indices}]}{%
		\ifthenelse{\equal{#1}{G}}{\item[\textbf{Generic formulation}]}{%
			\ifthenelse{\equal{#1}{S}}{\item[\textbf{Dispatch and Frequency regulation}]}{\item[\textbf{Real-time control}]}}}
}

\makenomenclature

\newcommand{\norm}[1]{\left\lVert#1\right\rVert}

\newcommand\Item[1][]{%
  \ifx\relax#1\relax  \item \else \item[#1] \fi
  \abovedisplayskip=0pt\abovedisplayshortskip=0pt~\vspace*{-\baselineskip}}

\newcommand{\babyblue}[1]{{\leavevmode\color{black}#1}}

\title{Control of Battery Storage Systems for the Simultaneous Provision of Multiple Services}
\author{Emil Namor, Fabrizio Sossan, Rachid Cherkaoui and Mario Paolone}

\author{%
\IEEEauthorblockN{Emil~Namor,~\IEEEmembership{Student~Member,~IEEE}, Fabrizio~Sossan,~\IEEEmembership{Member,~IEEE}, Rachid Cherkaoui, ~\IEEEmembership{Senior Member,~IEEE},  Mario~Paolone,~\IEEEmembership{Senior Member,~IEEE}.}

\thanks{The authors are with the Distributed Electrical Systems Laboratory, \'Ecole Polytechnique F\'ed\'erale de Lausanne, Switzerland (EPFL), e-mail: \{emil.namor, fabrizio.sossan, rachid.cherkaoui, mario.paolone\}@epfl.ch.

This work was supported in part by the Swiss Competence Center for Energy Research and in part by the European Union's Horizon 2020 research and innovation program under agreement no. 773406.

}
}

\begin{document}

\maketitle

\begin{abstract}
In this paper, we propose a control framework for a battery energy storage system to provide simultaneously multiple services to the electrical grid. The objective is to maximise the battery exploitation from these services in the presence of uncertainty (load, stochastic distributed generation, grid frequency). The framework is structured in two phases. In a period-ahead phase, we solve an optimization problem that allocates the battery power and energy budgets to the different services. In the subsequent real-time phase the control set-points for the deployment of such services are calculated separately and superimposed. The control framework is first formulated in a general way and then casted in the problem of providing dispatchability of a medium voltage feeder in conjunction to primary frequency control. The performance of the proposed framework are validated by simulations and real-scale experiments, performed with a grid-connected 560~kWh/720~kVA Li-ion battery energy storage system.
\end{abstract}

\section{Introduction} \label{sec:intro}

\subsection{Motivations}

Battery energy storage systems (BESSs) are a promising technology due to their inherent distributed nature, their ability to inject bidirectional power flows, their high power ramping and ability to provide a set of different grid services. 
As of today, BESSs are being deployed to provide several different services, such as peak shaving \cite{oudalov2007sizing}, energy management of microgrids \cite{belvedere2012microcontroller} and stochastic resources \cite{sossan2016achieving,ke2015control} and frequency and voltage regulation \cite{christakou2014primary,oudalov2007optimizing}. 
Such deployment is still slowed down by the high cost of these devices. While this cost is decreasing due to technological developments and economies of scale, a viable approach to optimize the exploitation of such devices is the development of control strategies able to provide simultaneously more than one of the services listed above. This allows for a better exploitation of the BESS from a technical and economical point of view.
More specifically, the simultaneous provision of multiple services via BESSs is of interest with respect to two aspects. First, different applications have different energy and power requirements. Some are ``energy intensive'', i.e. they need a large amount of energy but low instanteous power (e.g. peak shaving). Other are ``power intensive'', i.e. require higher levels of power but not high amount of energy (e.g. primary frequency regulation) \cite{hittinger2012properties}. Such different services could be coupled to match at best the energy and power ratings of the batteries. Second,  batteries are normally sized to provide a single service continuously. However, the actual daily \vphantom{exploitation} \textcolor{black}{deployment of power and exploitation of energy capacity} vary due to the uncertainty of the stochastic resources to which they are coupled \textcolor{black}{(e.g. uncontrollable loads and PV generation in \cite{sossan2016achieving}), or of the pricing signals that they track (e.g. energy and balancing power prices in \cite{wu2015energy})}. \vphantom{Therefore, the whole capacity of a battery is exploited quite rarely.} \textcolor{black}{Therefore, the deployment of such services rarely requires the exploitation of the whole BESS capacity.} When a portion of the \vphantom{battery}\textcolor{black}{BESS energy} capacity remains unexploited by \textcolor{black}{the deployment of} its main service, it could be allocated to a secondary service, to be deployed in parallel. In other words, coupling multiple services together may allow to exploit at best the batteries coupled with stochastic resources.

\babyblue{\subsection{Literature Survey}

The relevance of application sinergies for energy storage devices has been pointed out in general terms in \cite{eyer2010energy}. Several works, in the existing technical literature, propose approaches to provide simultaneously multiple grid services and demonstrate their effectiveness by simulations \cite{wasowicz2012evaluating,wu2015energy,kazemi2017long,kazemi2017operation,drury2011value,cheng2016co,megel2015scheduling,vrettos2013combined,shi2017using,xi2014stochastic,moreno2015milp,perez2016effect}. These references differ from each other for the kind of services they provide and how they account for BESS operational constraints. BESS services can be classified in 3 mainstream categories:
\begin{enumerate}
	\item energy arbitrage (EA), i.e. buying and selling electricity to generate a revenue;
	\item provision of ancillary services (AS). These are a set of services that batteries can provide to grid operators to enhance the system reliability (e.g. frequency response and regulation). The provision of these services is normally regulated by auction based systems and markets;
	\item achievement of control objectives for the local grid (i.e. local objectives (LO) ), like congestion management, voltage regulation at LV and MV level or self-consumption.
\end{enumerate}
The applications described in \cite{wasowicz2012evaluating,wu2015energy,kazemi2017long,kazemi2017operation,drury2011value,cheng2016co,megel2015scheduling,vrettos2013combined,shi2017using,xi2014stochastic,moreno2015milp,perez2016effect} are designed to provide combinations of the aforementioned services, as summarized in Table \ref{tab:literatureSurvey}.
\begin{table}
	\centering
	\caption{Recent literature on clustering of BESS applications in power systems}
	\label{tab:literatureSurvey}
	\begin{tabular}{| c | c |}
		\hline
		Services provided & References \\
		\hline
		EA + AS & \cite{wasowicz2012evaluating,wu2015energy,kazemi2017long,kazemi2017operation,drury2011value,cheng2016co} \\
		LO + AS & \cite{megel2015scheduling,vrettos2013combined,shi2017using} \\
		LO + EA + AS & \cite{xi2014stochastic,moreno2015milp,perez2016effect} \\
		\hline
	\end{tabular}
\end{table}

In such references, operation scheduling problems for energy storage systems considering multiple services are formulated. These aim at maximising the economic revenue generated for a standalone storage systems exploiting multiple revenue streams. This objective is sought in different pricing contexts and the common result is that by jointly providing multiple services, the BESS economic income is increased. Nonetheless, energy storage systems are often used in two further configurations \cite{kazemi2017operation}: \emph{i)} used by system operators to improve system reliability (e.g. \cite{nick2014optimal,zhao2015review}) or \emph{ii)} in conjunction with other resources such as distributed generation \cite{akhavan2014optimal}, flexible demand \cite{brahman2015optimal} or electric vehicles \cite{shafiee2012impacts}. 

Besides the objective of the proposed scheduling problems, the references listed in Table \ref{tab:literatureSurvey} focus on different aspects of the control framework needed to provide multiple services simultaneously. Several references propose specific methods for storage technologies other than BESSs: compressed air energy storage \cite{drury2011value}, fleets of thermostatically controlled loads \cite{vrettos2013combined}, or fleets of distributed BESSs \cite{megel2015scheduling}. References \cite{vrettos2013combined} and \cite{shi2017using}, besides the formulation of the scheduling problem, describe the real-time control to implement the proposed strategies. References \cite{kazemi2017long,kazemi2017operation,shi2017using} propose a robust optimization approach to deal with uncertainties related to price signals and reserve deployment. Finally \cite{kazemi2017long} analyses how providing multiple services simultaneously affects the BESS life time.

\subsection{Paper's contributions}

We consider the case of a BESS installed in a distribution feeder supplying uncontrollable loads and integrating a considerable amount of distributed generation. The scheduling problem of such BESS consists in allocating portions of its power and energy capacity to achieve different technical objectives, such as the dispatch of the active power demand of the feeder and the provision of primary frequency regulation power to the upper grid layer. 
Although the proposed framework can be adapted to maximise the revenue coming from providing difference ancillary services in a price-taking setting (as shown in Appendix~\ref{AN:1}), it is formulated with the objective of maximising the capacity of providing ancillary services. The reason for this is that the price taking assumption is not scalable with the number of units participating in the markets. In other words, if many units were to participate in the ancillary services market, an open-loop price signal would not be representative of their aggregated reaction. Recent works in \cite{mohsenian2016coordinated} and \cite{mohsenian2016optimal} addresses the problem of decision making for battery systems in a price-setting context, but they solely focus on energy arbitrage, whereas we consider multiple simultaneous services.

Specifically, we focus on the problem of jointly dispatching the operation of an active distribution feeder and provide primary frequency regulation. We provide first a formulation of a general control framework for the provision of multiple simultaneous grid services via BESSs, i.e. a formulation that is agnostic to the services that are provided. This solution does not require coordination mechanisms with other resources or with the upper grid layer nor an extensive communication infrastructure and can be considered as a bottom-up approach to augment the ability of BESSs to provide useful services to the grid. 
The proposed control has two time layers: (i) a period-ahead and (ii) a real-time one. In the first, we solve an optimization problem that allocates a power and an energy budgets to each considered service. This is done to maximize the exploitation of the BESS energy capacity and ensure continuous operation by managing the BESS stored energy. In the real-time stage, the power setpoints needed for each service are computed independently and superimposed. 
Based on such general framework, we describe then a BESS control scheme for dispatching the operation of a distribution feeder, such as in \cite{sossan2016achieving} and for primary frequency regulation. We show the performance of this control both in simulations and via experimental results obtained by implementing the proposed framework to control a grid-connected 560~kWh/720kVA BESS.
	The contributions of the paper, with respect to the existing literature are:
	\begin{itemize}
		\item the formulation of a complete algorithmic toolchain to control a BESS in order to provide multiple services simultaneously. This framework differs from the existing literature in: \emph{i)} the generic formulation of the scheduling problem, \emph{ii)} the technical rather than revenue-driven control objective, \emph{iii)} the consideration of the stochastic behaviour of the services deployment (due to the uncertainties in the forecast of the feeder prosumption as well as in the energy needed to perform PFR) and exploitation of robust optimization techniques to hedge against uncertainty and achieve reliable real-time operation (similarly to \cite{kazemi2017operation}). 
		\item the formulation of a control strategy to manage a BESS connected within a MV feeder, together with a set of heterogeneous resources (loads and PV generations), in order to dispatch the operation of the same feeder and exploit the remaining capacity to provide PFR.
		\item the experimental validation of the proposed control tool-chain, providing solid empirical evidences on the applicability, actionability, and performance of the proposed scheduling and control algorithms. 
		In the best of the Authors’ knowledge, this is the first work providing such experimental validation for a BESS control scheme considering multiple simultaneous services.		
	\end{itemize}
}

The paper is organised as follows. Section \ref{sec:genericPS} proposes the general formulation of the control problem of providing multiple services simultaneously via a BESS. Section \ref{sec:specificPS} casts the proposed framework in the specific context of the provision of power for dispatching the operation of an active distribution feeder and for primary frequency regulation (PFR). Section \ref{sec:results} presents results, obtained both via simulations and experiments, that validates the proposed framework. Finally, Section \ref{sec:conclusion} summarizes the original contributions and main outcomes of the paper and proposes directions for further research.

\section{Problem formulation} \label{sec:genericPS}
	
We consider the problem of scheduling the operation of a BESS with energy capacity $E_{nom}$ and maximum power $P_{max}$, for a time window $T$. During each time window, the BESS provides $J$ services, each denoted by the subscript $j=1,\dots,J$. \vphantom{Let $\mathcal{P}_j$ and $\mathcal{E}_j$ denote the power and energy budgets necessary to provide the service $j$ in the incoming time window $T$.}

\begin{figure}
	\begin{picture}(100,190)
	\put(0,0){\includegraphics[width=.48\textwidth]{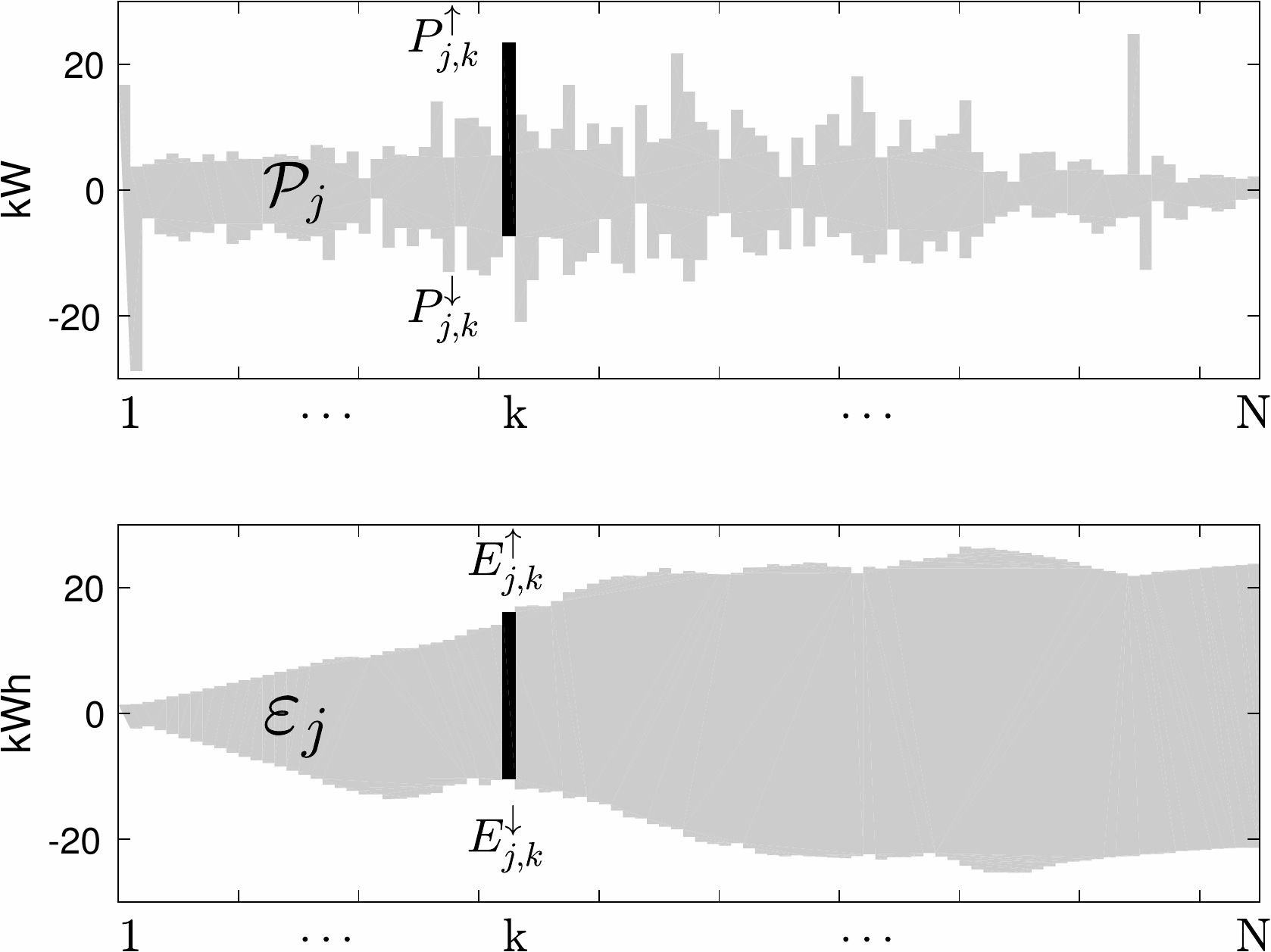}}
	\put(0,177){(a)}
	\put(0,75){(b)}
	\end{picture}
	\caption{Example of (a) power and (b) energy budgets for a service $j$.}
	\label{fig:budgetExamples}
\end{figure}

\textcolor{black}{Each service $j$ is characterized by an energy budget $\mathcal{E}_j$ and a power budget $\mathcal{P}_j$. These are the shares of the BESS energy capacity and power necessary along the time window $T$ to deploy the service $j$.}
\vphantom{$\mathcal{P}_j$ and $\mathcal{E}_j$ are computed before the time window of interest and are function of exogeneous input quantities denoted by $\theta$ (forecasts) and of a set of decision variables denoted by $x$. We formulate an optimization problem to determine the decision vector $x$ (and hence the power and energy budgets $\mathcal{P}_j$ and $\mathcal{E}_j$ for $j=1,\dots,J$) such that the exploitation of the BESS energy capacity is maximised.}
\textcolor{black}{The power and energy budgets $\mathcal{P}_j$ and $\mathcal{E}_j$ necessary for each service are functions of a set of tunable control parameters (composing the decision vector of the scheduling problem and hereafter denoted by $x$) as well as of variables modelling the uncertainty of the operating conditions related to each service (hereafter $\theta$). 
The dependency of $\mathcal{P}_j$ and $\mathcal{E}_j$ on $\theta$ is introduced to account for the fact that the deployment of the considered services need to be ensured in the occurrence of any scenario of their power demand (practical examples are provided in Section \ref{sec:specificPS}). We formulate an optimization problem to determine the value of decision vector $x$ (and hence the power and energy budgets $\mathcal{P}_j$ and $\mathcal{E}_j$ for $j=1,\dots,J$) that maximizes the portion of BESS energy capacity made available for the provision of the services in $J$.} 
We discretize the window of duration $T$ in $N$ time \vphantom{intervals} \textcolor{black}{steps} of duration $T/N$, each denoted by the subscript $k$, with $k = 1,\dots,N$. 
Formally, the power budget of the service $j$ \textcolor{black}{at time step $k$ is denoted $\mathcal{P}_{j,k}$ and} is defined as the interval of the expected power values that the service could require at $k$. These are between the \textcolor{black}{minimum and maximum}\vphantom{maximum negative and maximum positive} expected power realizations for that service, namely \textcolor{black}{in the interval} $\mathcal{P}_{j,k}=\left[ P_{j,k}^\downarrow,P_{j,k}^\uparrow\right]$. The power budget along a time period $T$ is defined as the sequence of such \textcolor{black}{intervals}:
\begin{align}
\mathcal{P}_j=\left\lbrace\left[ P_{j,k}^\downarrow(x,\theta),P_{j,k}^\uparrow(x,\theta)\right],k=1,\dots,N\right\rbrace.
\end{align}
Similarly, the application will require an energy budget
\begin{align}
\mathcal{E}_j=\{\left[ E_{j,k}^\downarrow(x,\theta),E_{j,k}^\uparrow(x,\theta)\right],k=1,\dots,N\}.
\end{align} 
An example of energy and power budets is reported in Fig. \ref{fig:budgetExamples}. 
The set of widths of such energy budget trajectory is defined as:
\begin{align}
w(\mathcal{E}_j(x,\theta)) & = \{E_{j,k}^\uparrow(x,\theta)-E_{j,k}^\downarrow(x,\theta),k=1,\dots,N\}.
\end{align}
Moreover, we define the operation of sum of budgets of different services (using energy budget as example) as\footnote{Eq. \eqref{eq:sum} can be interpreted as the computation of the overall energy budget within $T$ required by all services $J$.}:
\begin{align}
\begin{split}
&\sum_j \mathcal{E}_j(x,\theta) = \\ 
& = \left\lbrace\left[ \sum_{j=1}^JE_{j,k}^\downarrow(x,\theta),\sum_jE_{j,k}^\uparrow(x,\theta)\right],k=1,\dots,N\right\rbrace. \label{eq:sum}
\end{split}
\end{align}
The problem of providing multiple concurrent services with a BESS, while ensuring feasible operation can now be formulated in generic terms. We seek to maximise the set of widths of the energy budget resulting from the sum of the energy budgets $\mathcal{E}_j$ with $j=1,\dots,J$, within a given time window $T$, while respecting the BESS power and energy capabilities. The resulting decision problem is:
\begin{align}
& x^o = \underset{x}{\text{arg~max}} \norm{ w \left( \sum_{j=1}^J{\mathcal{E}_{j}(x,\theta)} \right) } \label{eq:generalFormulation}
\end{align}
subject to:
\begin{align}
& E_{init} + \sum_{j=1}^J{\mathcal{E}_{j}(x,\theta)} \in \left[E_{min},E_{max}\right] \label{eq:GF_energy} \\
& \sum_{j=1}^J{\mathcal{P}_j(x,\theta)} \in \left[ -P_{max},P_{max} \right] \label{eq:GF_power}
\end{align}
It is worth noting that it is possible to have a different objective function while exploiting the same framework presented here. In Appendix \ref{AN:1}, two variations seeking respectively the maximisation of the economical revenue and simple feasibility of operation are shown.
	
\section{Concurrent dispatch of a MV distribution feeder and primary frequency control} \label{sec:specificPS}

The scheme proposed in Section \ref{sec:genericPS} is now applied to control a BESS to dispatch of a MV distribution feeder and to provide PFR to the grid. We have observed that the battery capacity needed to dispatch a MV feeder as in \cite{sossan2016achieving} depends on the uncertainty of the forecast of the connected stochastic resources (loads and stochastic distributed generation). Whereas in some cases the battery capacity is barely sufficient to achieve this goal, in others a considerable portion of the battery capacity remains unutilized when the uncertainty of the prosumption forecast is small.

The choice of PFR as a second stacked service is because \emph{i)} large ramping duties of BESSs accomodate the increased demand for fast regulating power in power systems with a high penetration of  production from renewables and \emph{ii)} PFR is a ``power intensive'' application and is well-suited to be coupled with the dispatch service, which is instead ``energy intensive''.

\subsection{Day-ahead problem formulation}

\begin{figure}
  \includestandalone[mode=buildnew,width=.5\textwidth]{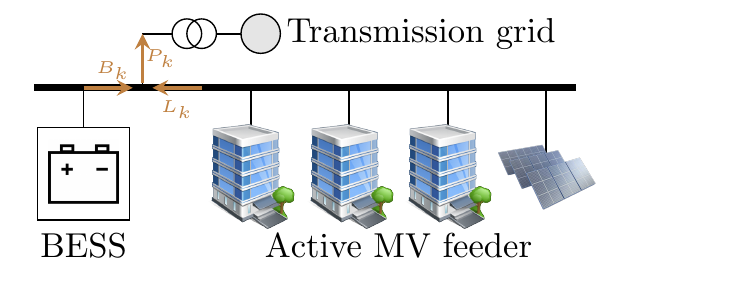}
\caption{Schematic of the experimental setup. The notation of the power flows refers to the real-time control described in section \ref{ssec:realtime}.}
\label{fig:feederSchematic}
\end{figure}

We want to operate a grid-connected BESS to dispatch the active power flow of a MV distribution system with heterogeneous resources, as in \cite{sossan2016achieving}, while providing also primary frequency regulation to the grid. Figure \ref{fig:feederSchematic} shows the main features of this setup.
The operation is performed over a T=24 hour period and planned every day for the next calendar day. Following the formulation presented in Section \ref{sec:genericPS}, we first define the power and energy budgets for the dispatch and PFR, namely $\mathcal{P}_D$, $\mathcal{P}_{FR}$, $\mathcal{E}_D$ and $\mathcal{E}_{FR}$. Based on these budgets, we formulate an optimization problem as in \eqref{eq:generalFormulation}-\eqref{eq:GF_power}.

\subsubsection{Power and energy budgets}
The dispatch service requires the battery to compensate for the mismatch between the aggregated prosumers power flow (denoted by $\mathbf{L}_k=L_1,\dots,L_N$) and a pre-established dispatch plan $\hat{\mathbf{P}}_k=\hat{P}_1,\dots,\hat{P}_N$, defined at 5-minutes resolution. The dispatch plan is the sum of two terms: the forecasted power profile of the feeder prosumption, $\hat{\mathbf{L}}=\hat{L}_1,\dots,\hat{L}_N$ and an offset power profile, $\mathbf{F}=F_1,\dots,F_N$, computed to keep the BESS stored energy within proper limits:
\begin{align}
\hat{P}_k = \hat{L}_k + F_k \text{~~~for~~~} k=1,\dots,N
\end{align} 
We obtain, with a forecasting tool from the literature \cite{sossan2016achieving}, the daily forecasted profile of the feeder prosumption as well as the deviations from the forecasted profile in the highest and lowest demand scenarios, denoted by $\mathbf{L}^\uparrow=L^\uparrow_1,\dots,L^\uparrow_N$ and $\mathbf{L}^\downarrow=L^\downarrow_1,\dots,L^\downarrow_N$.
The maximum \vphantom{and minimum} positive and negative BESS power requirements for the dispatch service are therefore defined as the sum over $k$ of the offset power $F_k$ and of $L^\uparrow_k$ and $L^\downarrow_k$, respectively. \textcolor{black}{With respect to the general definitions of $x$ and $\theta$ given Section \ref{sec:genericPS}, the terms $L^\uparrow_k$ and $L^\downarrow_k$ are input quantities (i.e. $\lbrace \textbf{L}^\uparrow, \textbf{L}^\downarrow \rbrace$ are in $\theta$) whereas the offset power $\mathbf{F}$ is a decision variable, determined by the optimization problem defined hereafter (i.e. $\mathbf{F}$ is in $x$).} The power budget is therefore defined as:
\begin{align}
\begin{split}
\mathcal{P}_D & = \{\left[ P_{D,k}^\downarrow(x,\theta),P_{D,k}^\uparrow(x,\theta)\right],k=1,\dots,N\} \\
& = \{\left[ F_k + L_k^\downarrow, F_k + L_k^\uparrow \right],k=1,\dots,N\} \label{eq:dispPB}
\end{split}
\end{align}
The associated energy budget is:
\begin{align}
\begin{split}
\mathcal{E}_D & = \{\left[ E_{D,k}^\downarrow(x,\theta),E_{D,k}^\uparrow(x,\theta)\right],k=1,\dots,N\} \\
& = \left[ \frac{T}{N} \sum_{i=1}^k (F_i+L_i^\downarrow) , \frac{T}{N} \sum_{i=1}^k (F_i+L_i^\uparrow)\right] \label{eq:dispEB}
\end{split}
\end{align}
with $k=1,\dots,N$.

The primary frequency regulation service requires the battery to provide a power proportional to the deviation of the frequency from its nominal value $\Delta f_k = f_k - f_n$ \textcolor{black}{\cite{xu2014bess}}, with a \textcolor{black}{proportionality coefficient hereafter denoted by $\alpha$:}
\begin{align}
	\color{black} P_{FR,k} = \alpha \Delta f_k = \alpha \left( f_k - f_n\right).  
\end{align}
\textcolor{black}{The unit of measurement of $\alpha$ is kW/Hz.}\vphantom{ and it defines the PFR regulating power. The coefficient $\alpha$ (in kW/Hz) determinates the PFR regulating power.} The \textcolor{black}{instantaneous} requested power cannot be forecasted since frequency deviations are difficult to predict. Therefore, the power budget required by this application will correspond to a constant profile, equal to the maximum power that frequency regulation may require. Since grid codes typically require complete activation of primary reserves for frequency deviations of more than $\Delta f_{max}=$ 200~mHz \cite{xu2014bess}, the power budget can be defined as:
\begin{align}
\begin{split}
\mathcal{P}_{FR} & = \{\left[ P_{FR,k}^\downarrow(x,\theta),P_{FR,k}^\uparrow(x,\theta)\right],k=1,\dots,N\}\\
& = \left[ -0.2 \alpha \cdot \mathbf{1}, 0.2 \alpha \cdot \mathbf{1}\right] \label{eq:pfrPB}
\end{split}
\end{align}
Where $\mathbf{1}$ is the all-one vector of length $N$. The energy budget necessary to ensure feasible operation for this service within a given time interval can be inferred statistically. In particular, we examined grid frequency data of the European grid from the last 2 years. Data have been collected by a PMU-based metering system installed on the EPFL campus \cite{pignati2015real}.
Since frequency regulation requires the injection of a power $P_k=\alpha\Delta f_k$, the energy required by the grid during a given time window $T$ is:
\begin{align}
\begin{split}
\color{black} E_{FR,k} & = \frac{T}{N} \sum_{i=0}^k P_{FR,i} = \frac{T}{N} \sum_{i=0}^k (\alpha \Delta f_i) \\ 
& = \alpha \left( \frac{T}{N} \sum_{i=0}^k \Delta f_i \right) = \alpha W_{f,k}
\end{split}
\end{align}
for $k=1,\dots,N$ and where $W_{f,k}$ denotes the integral of frequency deviations over a period of time and it is to be interpreted as the \textit{energy content} of the signal given by the frequency deviation from its nominal value. The upper and lower bounds for $W_{f,k}$ for $k=1,\dots,N$ can be inferred from a statistical analysis of historical frequency deviation time-series (reported in Appendix \ref{appendix:Wf}). These are defined hereafter as $\mathbf{W}_f^\uparrow = W_{f,1}^\uparrow,\dots,W_{f,N}^\uparrow$ and $\mathbf{W}_f^\downarrow = W_{f,1}^\downarrow,\dots,W_{f,N}^\downarrow$. \textcolor{black}{With regard to the general definitions of $x$ and $\theta$ given in Section \ref{sec:genericPS}, the terms $W^\uparrow_k$, $W^\downarrow_k$ (as well as $\Delta f_{max}$ in \eqref{eq:pfrPB}) are input quantities (i.e. $\lbrace \textbf{W}^\uparrow, \textbf{W}^\downarrow, \Delta f_{max} \rbrace$ are in $\theta$) whereas $\alpha$ is a decision variable, determined by the optimization problem defined hereafter (i.e. $\alpha$ is in $x$).} The energy budget for frequency regulation is then defined as:
\begin{align}
\begin{split}
\mathcal{E}_{FR} & = \{\left[ E_{FR,k}^\downarrow(x,\theta),E_{FR,k}^\uparrow(x,\theta)\right],k=1,\dots,N\} \\
& = \{\left[ \alpha W_{f,k}^\downarrow, \alpha W_{f,k}^\uparrow \right],k=1,\dots,N\} \label{eq:pfrEB}
\end{split}
\end{align}

\subsubsection{Decision problem formulation} \label{sssec:specificdecision}
relying on the definitions given in Section \ref{sec:genericPS}, it is $x = [\alpha, \mathbf{F}]$ and $\theta = [\Delta f_{max}, \mathbf{W}_f^\downarrow, \mathbf{W}_f^\uparrow, \mathbf{L}^\downarrow, \mathbf{L}^\uparrow]$ and the objective function in \eqref{eq:generalFormulation}, corresponds therefore to:
\begin{align}
\begin{split}
& w \left( \sum_j \mathcal{E}_j \right) = w \left( \mathcal{E}_D + \mathcal{E}_{FR}\right) \\% = w(\mathcal{E}_D) + w(\mathcal{E}_{FR})\\
& = \left( \frac{T}{N} \sum_{i=0}^k (F_i + L_i^\uparrow) +\alpha W_{f,k}^\uparrow \right) + \\ & - \left( \frac{T}{N} \sum_{i=0}^k (F_i + L_i^\downarrow) + \alpha W_{f,k}^\downarrow\right)\\
& = \left( \frac{T}{N} \sum_{i=0}^k (L_i^\uparrow) - \frac{T}{N} \sum_{i=0}^k (L_i^\downarrow) \right) + \alpha \left( W_{f,k}^\uparrow - W_{f,k}^\downarrow\right) \\
& \text{with } k = 1,\dots,N. \label{eq:EBsum}
\end{split}
\end{align}
Since $\alpha$ is the only control variable in the expression above, the objective to maximize $w \left( \sum_j \mathcal{E}_j \right)$ in \eqref{eq:generalFormulation} reduces to maximizing $\alpha$, subject to \eqref{eq:GF_energy}\eqref{eq:GF_power}. The problem \eqref{eq:generalFormulation}-\eqref{eq:GF_power} is as:
\begin{align}
& \left[ \alpha^o, \mathbf{F}^o \right] = \underset{\alpha \in \mathbb{R}^+, \mathbf{F} \in \mathbb{R}^N}{\text{arg~max}} \left( \alpha \right) \label{eq:specificFormulation}
\end{align}
subject to:
\begin{align}
& E_{init} + \mathcal{E}_{D}(x,\theta) + \mathcal{E}_{FR}(x,\theta) \in \left[E_{min},E_{max}\right] \label{eq:SFenergy}\\
& \mathcal{P}_{D}(x,\theta) + \mathcal{P}_{FR}(x,\theta) \in \left[ -P_{max},P_{max} \right] \label{eq:SFpower}
\end{align}
By expressing explicitly the dependency of the power and energy budgets on the parameters and control variables, the problem \eqref{eq:specificFormulation}-\eqref{eq:SFpower} becomes:
\begin{align}
& \left[ \alpha^o, \mathbf{F}^o \right] = \underset{\alpha \in \mathbb{R}^+, \mathbf{F} \in \mathbb{R}^N}{\text{arg~max}} \left( \alpha \right) \label{eq:specificFormulation_2}
\end{align}
subject to:
\begin{align}
& E_{init} + \frac{T}{N}\sum_{i=1}^k \left( F_i + L^\uparrow_i \right) + \alpha W_{f,k}^\uparrow \le E_{max} \label{eq:cb1} \\
& E_{init} + \frac{T}{N}\sum_{i=1}^k \left( F_i + L^\downarrow_i \right) + \alpha W_{f,k}^\downarrow \ge E_{min} \label{eq:cb2} \\
& F_k + L^\uparrow_k + 0.2 \alpha \ge P_{max} \\
& F_k + L^\downarrow_k + 0.2 \alpha \ge -P_{max} \label{eq:specificFormulation_2_end}
\end{align}
with $k=1,\dots,N$.

\subsubsection{Determination of $E_{min}$ to include the BESS efficiency} \label{sssec:eta}

The notion of battery round-trip efficiency is incorporated in the decision problem \eqref{eq:specificFormulation_2}-\eqref{eq:specificFormulation_2_end} with an empirical  two-stage approach by enforcing conservative limits for the battery stored energy. This process is explained in the following. First, the problem \eqref{eq:specificFormulation_2}-\eqref{eq:specificFormulation_2_end} is solved implementing the nominal battery state-of-energy limits (i.e. $E_{max}=E_{nom}$ and $E_{min}=0$). Second, the following finite impulse response model \cite{sossan2016achieving,fortenbacher2017optimal}:
\begin{align}
E_k = E_0 + \frac{T}{N} \sum_{i=1}^k \eta_i B_i, && \eta_i = 
\begin{cases}
\beta & B_i \ge 0\\
1/\beta &  B_i < 0
\end{cases},
\end{align}
where $B_i$ is the total power injected or absorbed by the BESS at time $i$ and $\eta_i$ the BESS efficiency, is used to model the stored energy $E_k$ of a non ideal BESS for the set of simulation scenarios presented in Section~\ref{ssec:sim}. The energy stored at the end of each day in a BESS modeled as ideal ($\eta=1$) and non ideal ($\eta=0.96$\footnote{the value of $\eta=0.96$ has been determined experimentally for the 560~kWh/720~kVA BESS used in this work.}) are compared and the largest difference over the all set of simulations is used to impose a conservative bound to the minimum stored energy constraint \eqref{eq:cb2}. For example, in the case proposed in Section~\ref{ssec:sim}, the largest difference is 4\% of $E_{nom}$, therefore we adopt $E_{min}=0.05E_{nom}$.
It is worth noting that this approach allows to define the energy budgets independently for each service and sum them as in \eqref{eq:EBsum}. In other words, it achieves a separation of concerns between services, which can be designed independently from each other and stacked together at the end of the process. Also, it is worth noting that the round-trip efficiency of modern Li-ion based BESS is generally above 90\% \cite{qian2011high,castillo2014grid,fortenbacher2014modeling}.
An accurate investigation of the modelling errors, considering also less efficient storage technologies (like fuel cells), is postponed to future works.

\subsection{Real-time control} \label{ssec:realtime}
\begin{figure}
  \centering
  \includestandalone[mode=buildnew,width=.5\textwidth]{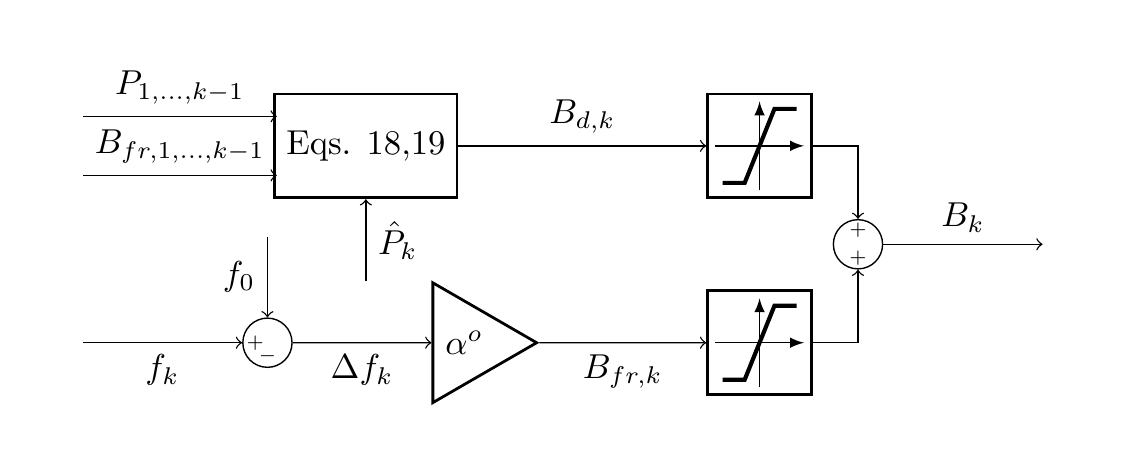}
  \caption{Scheme of the BESS real-time control}
  \label{fig:realtime}
\end{figure}

The proposed algorithm consists in solving a planning problem for the next calendar day of operation, determining the values of \textcolor{black}{the coefficient $\alpha^o$} and of the offset profile $\textbf{F}^o$ and in a real-time control problem. The latter is not the main contribution of the present work, however it is summarized hereafter and illustrated in Figure \ref{fig:realtime} for the sake of clarity. 
The real-time control determines the battery active power setpoint $B_{k}$ with 1-second resolution. In the following, the index $k$ denotes the 1-second resolution time interval. $B_k$ is the algebraic sum of the setpoints $B_{d,k}$ and $B_{fr,k}$ determined respectively for the dispatch and the PFR by two independent control loops:
\begin{align}
B_{k}=B_{d,k}+B_{fr,k}.
\end{align}
The power setpoint $B_{d,k}$ is to compensate the tracking error $\epsilon_k$, which is the difference between the objectve feeder power $\hat{P}_k$ (from the dispatch plan, with 5 minutes resolution) and the mean deviation from this value within the 5 minutes interval. This deviation is the sum of two terms. The first is the mean of the feeder power measurements $P_i$ in the instants from the beginning of the current 5-minutes period and present, filtered out of the power requests due to the PFR, $B_{fr,i}$. The second is a short-term forecast of the load $\hat{L}_i$ over the remaining five minutes interval:
\begin{align}
\epsilon_k = \hat{P}_k - \frac{1}{300}\left( \sum_{i=0}^{k-1} (P_i-B_{fr,i}) + \sum_{i=k}^{\text{5~min}}\hat{L}_i\right).
\end{align}
The expression above is an energy objective over a 5~minutes horizon and the power setpoint to respect it is therefore defined as:
\begin{align}
B_{d,k} = \frac{1}{300-k} \cdot \epsilon_k.
\end{align}
The power setpoint for the frequency regulation $B_{fr,k}$ is calulated as:
\begin{align}
B_{fr,k} = \alpha^o \cdot \left( f_k - f_n \right).
\end{align}
In order to comply with the constraints imposed by the day-ahead policy, both setpoints are constrained within saturation tresholds, which are, notably, equal to $ \pm 0.2 \alpha^o$ for $B_{fr,k}$ and $ \pm (P_{max} - 0.2 \alpha^o)$ for $B_{d,k}$. The latter threshold is set such that the dispatch can require, istantaneously, all the power not reserved by the frequency regulation. It remains, nevertheless, that the dispatch power averaged over a 5 minutes period is expected to remain between $\mathbf{L}^\uparrow+\mathbf{F}^o$ or $\mathbf{L}^\downarrow+\mathbf{F}^o$. 

\begin{figure*}
\centering
\includegraphics[width=\textwidth,height=4cm]{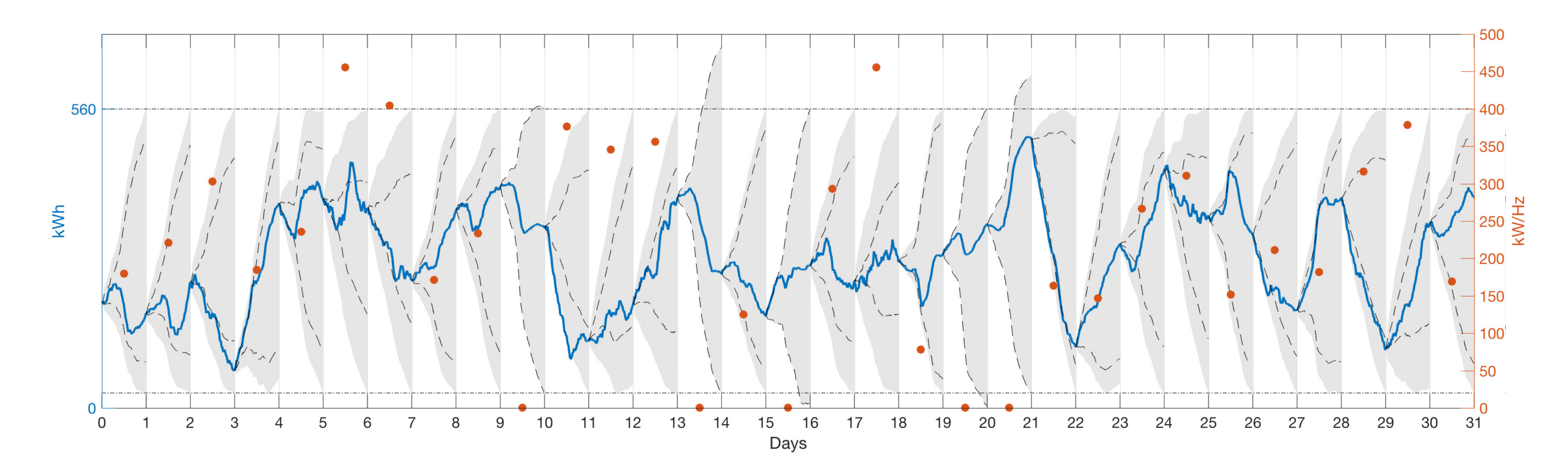}
\caption{Simulation results of 31 consecutive days of operation. Blue line: BESS stored energy; Grey area: total daily energy budget $\mathcal{E}_{D}+\mathcal{E}_{FR}$; Black dashed lines: bounds of the daily energy budget reserved to the dispatching service $\mathcal{E}_D$; red dots: daily values of $\alpha^o$ (referred to the right-hand y-axis).}
\label{fig:simsoe}
\end{figure*}

\section{Results} \label{sec:results}

The proposed planning and control strategy has been validated by simulations and experiments in a real-life grid. 

The goal of this validation effort is double. The simulations demonstrate the effectiveness of the proposed control architecture in the determination of the coefficient $\alpha^o$ and of the offset profile $\mathbf{F^o}$. The values found for such quantities allow to maximise the battery exploitation, while respecting the battery operational limits and therefore allowing for the continuous operation for a month. The experimental results validate the assumptions made in the control design and in the simulations and demonstrate the practical relevance and deployability of the proposed control architecture.

Both simulations and experiments are based on a setup with a 560~kWh/720~kVA Lithium-ion BESS installed at the EPFL campus in Lausanne, Switzerland, and connected to a 20~kV medium voltage feeder. The feeder interfaces 5 office buildings (300~kW global peak demand) and rooftop PV installations (90~kWp). Both historical data used in the simulations and real-time measurements of the power flows and grid frequency are obtained via a PMU-based metering system \cite{pignati2015real}. 

\subsection{Simulations} \label{ssec:sim}

Thirty-one consecutive days of operation are simulated. These 31 days are characterised by different initial SOE values\footnote{The SOE is here defined as the amount of stored energy normalized over the BESS nominal energy capacity $E_{nom}$.}, ranging from 12\% to 90\%, and determined by the operation of the previous days (the first day of the simulation the initial SOE has been set to 35\%).

Figure \ref{fig:simsoe} reports the profile of the energy stored the battery along the 31 days and the \textcolor{black}{daily energy budget for the dispatching service $\mathcal{E}_D$ and the} total daily energy budget ($\mathcal{E}_D+\mathcal{E}_{FR}$), calculated as a function the stochastic forecasting model of the demand and frequency (i.e. on the basis of $[\mathbf{L}^\uparrow, \mathbf{L}^\downarrow, \mathbf{W}_{f}^\uparrow, \mathbf{W}_{f}^\downarrow]$). Figure \ref{fig:simsoe} shows as well the values assumed daily by $\alpha^o$. It can be observed that the total daily energy budgets (grey areas) hit the BESS operational limits (SOE=5\% and SOE=100\%) in all days except for day 10, 14, 16, 20 and 21. This denotes that the day-ahead planning problem is able to schedule efficiently the offset profile $\mathbf{F}^o$ and the value $\alpha^o$ to exploit the full battery energy capacity accounting for the stochastic behaviour of frequency and demand. On the other hand, in the five days mentioned above, the grey area exceeds the SOE limits. This is because the uncertainity related to the demand (reflected by the sequences $\mathbf{L}^\uparrow$ and $\mathbf{L}^\downarrow$) prevents the feasibility of problem \eqref{eq:specificFormulation}-\eqref{eq:SFpower}. In such days, the solution of \eqref{eq:specificFormulation}-\eqref{eq:SFpower} provides an $\alpha^o$ equal to zero, i.e. no frequency regulation is performed. In all cases, the activated constraint in the solution of \eqref{eq:specificFormulation}-\eqref{eq:SFpower} has been the one on the energy budget sum.

Quantitative results from the simulations are collected in Table \ref{table:tab:simresults}: $SOE_0$ is the daily initial SOE in percentage, $\alpha^o$ the daily \textcolor{black}{coefficient for PFR} in kW/Hz, $F_{avg}$ the mean value of the offset profile and $\Delta SOE$ the overall SOE variation during the day due to the simultaneous deployment of the two services. Table \ref{table:tab:simresults} shows \textcolor{black}{the average, maximum and minimum values of such quantities over the 31 days simulation period. 
The average daily value of $\alpha^o$ is of 216.6~kW/Hz. This corresponds to the provision of up to 43~kW for PFR (considering $\Delta f_{max}$~=~200~mHz). In comparison to the work by the same Authors in \cite{sossan2016achieving}, where the control of the BESS aims exclusively at dispatching the operation of a MV feeder, we are able to provide power both for the dispatch and for PFR, while still ensuring the respect of the BESS operational constraints. This is done by taking advantage of the BESS capacity that remains unexploited by the dispatching operation, due to the daily variation of the uncertainty set of the prosumption defined by $\left[ L_k^\downarrow, L_k^\uparrow \right]$ for $k=1,\dots,N$. The black dashed lines in Figure \ref{fig:simsoe} delimit the energy budget reserved to the dispatching service $\mathcal{E}_D$. The width of this budget in days characterized by low uncertainty in the feeder prosumption forecast (e.g. days 5 or 17) is rather narrow and the unexploited battery capacity is therefore allocated to provide PFR (a high value of $\alpha^o$ is found). In days in which such uncertainty is high (e.g. days 18 to 20) almost all (or more than all) the battery capacity is needed to perform the dispatch, resulting in a very wide $\mathcal{E}_D$ and in a very low value of $\alpha^o$.}
\begin{table}
\centering
\caption{Simulation results}
\begin{tabular}{|c|c|c|c|c|c|}
\hline
& $SOE_0$ & $\alpha^o$ & $F_{avg}$ & $SOE_{min}|_T$ & $SOE_{max}|_T$ \\
& {[}\%] & [kW/Hz] & [kW] & [\%] & [\%] \\
\hline
Mean & 50.8 & 216.6 & 0.5 & 37.4 & 64.9 \\
Max & 90.3 & 455.7 & 10.0 & 61.6 & 90.7 \\
Min & 12.5 & 0.0 & -9.3 & 12.4 & 36.0 \\
\hline
\end{tabular}
\label{table:tab:simresults}
\end{table}
\begin{figure*}
\centering
\includegraphics[width=.45\textwidth]{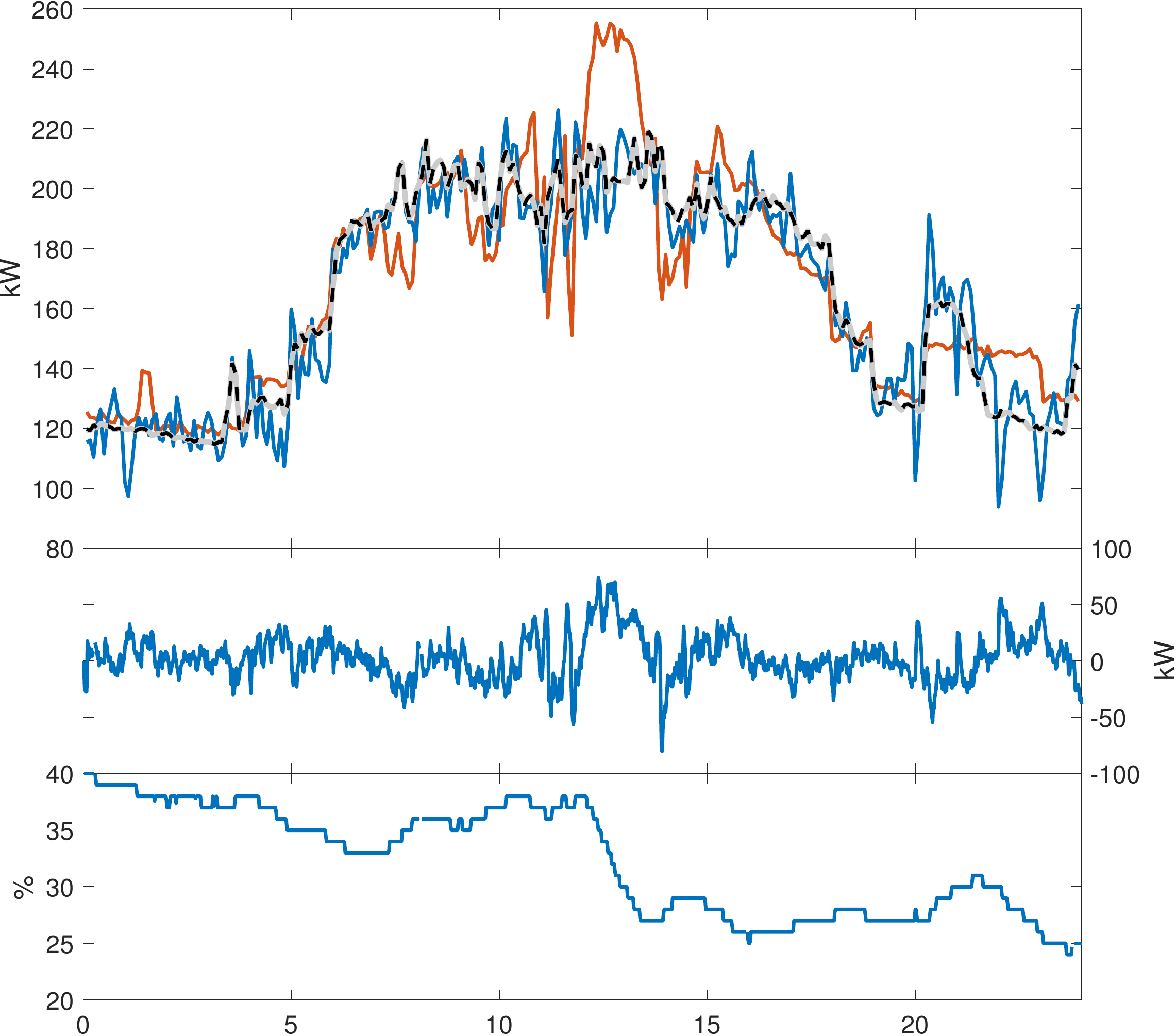}
\includegraphics[width=.45\textwidth]{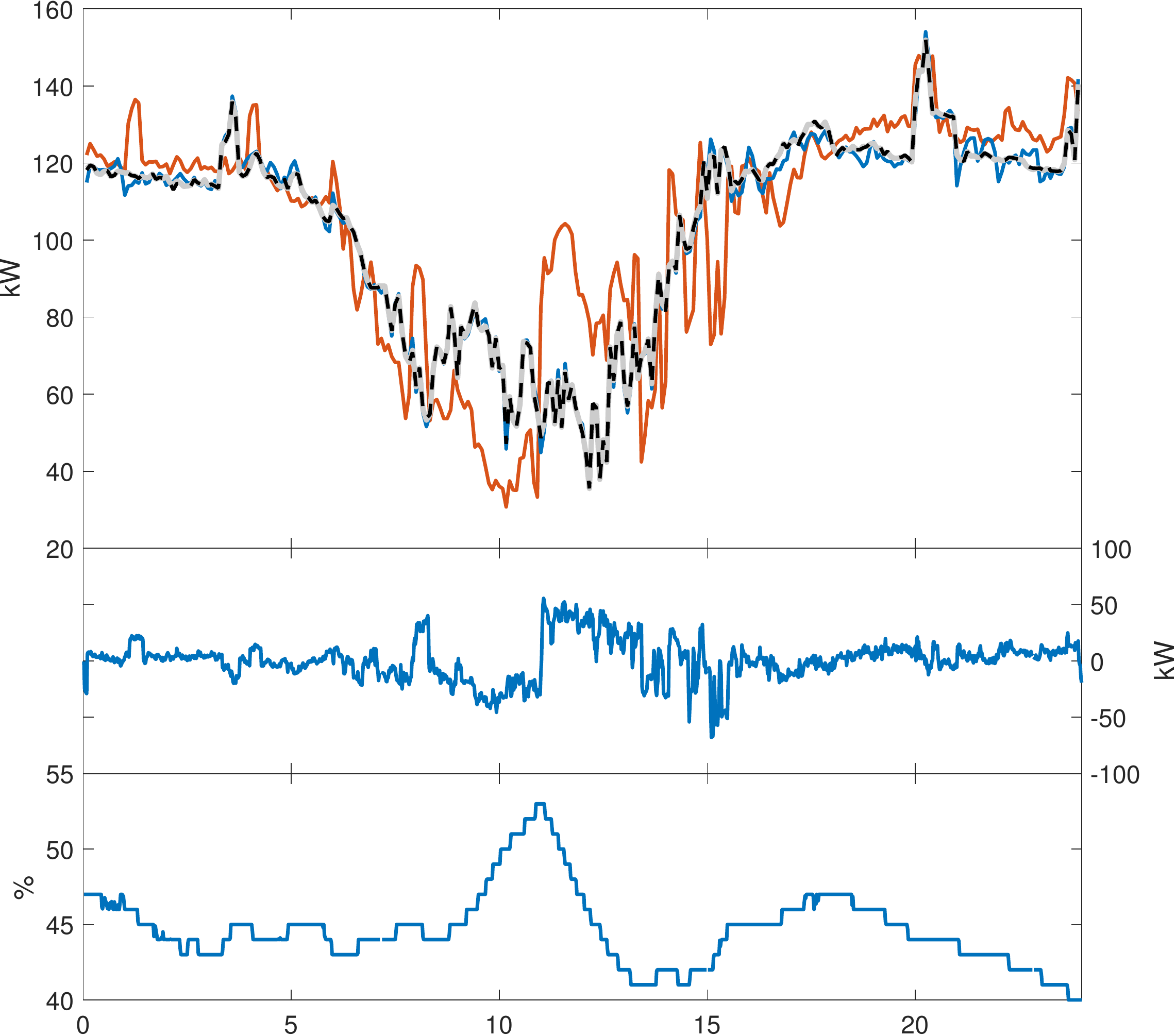}
\caption{Experimental results, \textbf{left: }day~1, \textbf{right: }day~2. \textbf{Upper plots -} feeder power profiles. Thick grey line: dispatch plan, red line: feeder prosumption, dashed black line: feeder real power (excluded the PFR power injection), blue line: feeder real power (with the PFR). \textbf{Middle plots -} BESS power injection. \textbf{Lower plots -} BESS SOE evolution.}
\label{fig:experimental_1}
\end{figure*}

\subsection{Experimental validation}

The described algorithm has been implemented in the controller of the 560~kWh/720~kVA Lithium-ion BESS. The results of 2 days of experiments are reported in this section.

Figure \ref{fig:experimental_1} shows the power and SOE profiles for two days of operation, an intra-week day and a weekend day (hereafter referred to as Day~1 and Day~2). \textcolor{black}{Numerical results are summarized in Table \ref{table:tab:expResults} and Table \ref{tab:exp_dispatch}.}
In Day~1, the day-ahead optimization procedure has determined a \textcolor{black}{value of $\alpha^o$} of 584~kW/Hz and an offset power of 0.84~kW on average. In Day~2, the \textcolor{black}{$\alpha^o$} has been found equal to 127~kW/Hz and the average offset power equal to -0.56~kW. 
\textcolor{black}{These values of $\alpha^o$ allow to exploit a portion of the battery capacity that would remain unexploited when providing power only to dispatch the operation of the MV feeder, as in\cite{sossan2016achieving}. In this case, the maximum amplitude of the energy budget that needs to be reserved for the dispatch, calculated as in \eqref{eq:dispEB} on the basis of the upper and lower worst case scenarios of the feeder prosumption ($L_k^\downarrow$ and $L_k^\uparrow$, with $k=1,\dots,N$), is of about the 54\% of the BESS nominal capacity for Day~1 and of about the 10\% for Day~2. The remaining capacity is fully exploited by the PFR application, thanks to the computation of a proper value of $\alpha^o$, by means of \eqref{eq:specificFormulation}-\eqref{eq:SFpower}.}

Table \ref{tab:exp_dispatch} collects the relevant metrics to evaluate the performance of the dispatch application when performed in conjunction with frequency regulation, i.e. the mean, RMS and maximum absolute values of the tracking error in these two days. The RMS value of the tracking error is about 0.5~kW over a feeder prosumption of about 130~kW on average.

\begin{table}
\centering
\caption{Experimental Results}
\begin{tabular}{|c|c|c|c|c|c|}
\hline
 & $SOE_0$ & $\alpha^o$ & $F_{avg}$ & $SOE_{min}|_T$ & $SOE_{max}|_T$ \\
& [\%] & [kW/Hz] & [kW] & [\%] & [\%] \\
\hline
Day 1 & 40 & 584 & 0.84 & 24 & 40 \\
Day 2 & 47 & 127 & -0.56 & 40 & 53 \\
\hline
\end{tabular}
\label{table:tab:expResults}
\end{table}

\begin{table}
\centering
\caption{Dispatch performance metrics (in kW)}
\begin{tabular}{|c|c|c|c|}
\hline
 & $\epsilon_{mean}$ & $\epsilon_{rms}$ & $\epsilon_{max}$ \\
\hline
Day 1 & -0.03 & 0.52 & 4.45 \\
Day 2 & 0.02 & 0.5 & 6.83 \\
\hline
\end{tabular}
\label{tab:exp_dispatch}
\end{table}

We note that, in both these two days, the energy demand for the two applications has been of opposite sign. For istance, in Day~1 the daily energy requested for the dispatch operation is of about 89~kWh, whereas the average power requested for the frequency regulation is of $-24$~kWh. The simultaneous deployment of these two services in this case generates a SOE drift that is lower than the one the dispatch alone would generate. It is worth noting that, when simultaneously providing multiple services, the saturation (or depletion) of the battery energy capacity would occur only if the power requests of all services corresponded to the upper (or lower) bounds of their budgets. If the uncertain processes related to the services are uncorrelated, as in the case of the dispatch and frequency regulation, the occurrence of this condition is reduced. Providing multiple services simultaneously, in this regard, may ensure more reliable operation, in the sense that failure due to complete depletion or saturation of the battery capacity would be less likely to occur. The downside of this is of course that an eventual failure would be more deleterious since multiple services would stop at once. This could be addressed by implementing strategies to prioritize the services in contingency situations, e.g. by selecting, before hitting the operational limits, which service is to drop and which to maintain.

\section{Conclusion} \label{sec:conclusion}

We have proposed an algorithm to schedule and control the operation of a battery energy storage system to provide multiple services simultaneously. Its objective is maximising the battery capacity exploitation in the presence of variable and stochastic energy and power requirements.

The proposed control consists in two phases. First, in the operation-scheduling phase the portion of battery power and energy capability to be allocated for each service is determined. This is accomplished by an optimization that takes into account the uncertainty in the forecasted power and energy requirements of each service. Second, in the real-time phase the different services are deployed by injecting in the grid a real power corresponding to the sum of the power setpoints of the individual services. 

The algorithm is first formulated in generic terms and then casted to the case of providing BESS power to simultaneously dispatch the active power flow of a distribution network and provide primary frequency regulation to the grid. For these two services the power and energy budgets are modelled in the planning problem by predictions delivered by forecasting tools. The solution of the operation-scheduling optimization problem provide, on a daily basis, \vphantom{the droop coefficient that allows for the maximum frequency regulating capacity}\textcolor{black}{the maximum value of the PFR regulating power that can be deployed} while respecting the battery operational constraints. It provides moreover the offset profile, i.e. the power needed, on a daily basis to restore the stored energy to a level that ensures continuous operation. 

The proposed control scheme is validated by simulations and experimentally. Simulations are obtained by applying the proposed scheme to a set of load and frequency data measured on-site and corresponding to one month of operation. Simulation results show that the proposed scheme does ensure continuous operation and does determine the maximum possible frequency regulating power that can be provided in conjunction to the dispatch application. Experiments are performed on a real-life grid by using a grid-connected 560~kWh/720~kVA lithium titanate BESS, connected to a medium voltage grid interfacing a set of office buildings and PV generating units. Results from 2 days of operations are shown and demonstrate the deployability of the proposed control scheme. In these two days of operation, a regulating power up to 117 and 25~kW respectively can be provided on top of the dispatch operation. The latter is performed with a RMS tracking error of about 0.5~kW.

Future works concern the development of contingency strategies to prioritize the services if the battery reaches its operational limits and an evaluation of the proposed control scheme applied to time horizons of different duration (e.g. intra-day, hourly operation).

\appendices

\section{Economic optimization and feasibility problems} \label{AN:1}

The objective of the cost function \eqref{eq:generalFormulation} is maximising the battery energy capacity exploited during a period of operation T. The same framework can be exploited to optimise the BESS operation considering different objectives. For instance, one could seek the value of $x$ that maximises the economical benefit of providing multiple concurrent services via an optimization function such as:
\begin{align}
& x^o = \underset{x}{\text{arg~max}} \sum{r_{j}}
\end{align}
subject to \eqref{eq:GF_energy}, \eqref{eq:GF_power} and:
\begin{align}
& r_{j} = f_j(\mathcal{E}_j,\mathcal{P}_j)
\end{align}
where $r_{j}$ is the revenue that the application $j$ can generate in period T, and is a function of the energy and power budgets reserved for that service.
Similarly, if the objective is simply to find a value for $x$ that ensures feasible operation, one could write:
\begin{align}
& x^o = \underset{x}{\text{arg~max}} \quad 1
\end{align}
subject to \eqref{eq:GF_energy} \eqref{eq:GF_power}.

\section{Computation of BESS energy needs for PFR} \label{appendix:Wf}

\begin{figure}
\includegraphics[width=.5\textwidth]{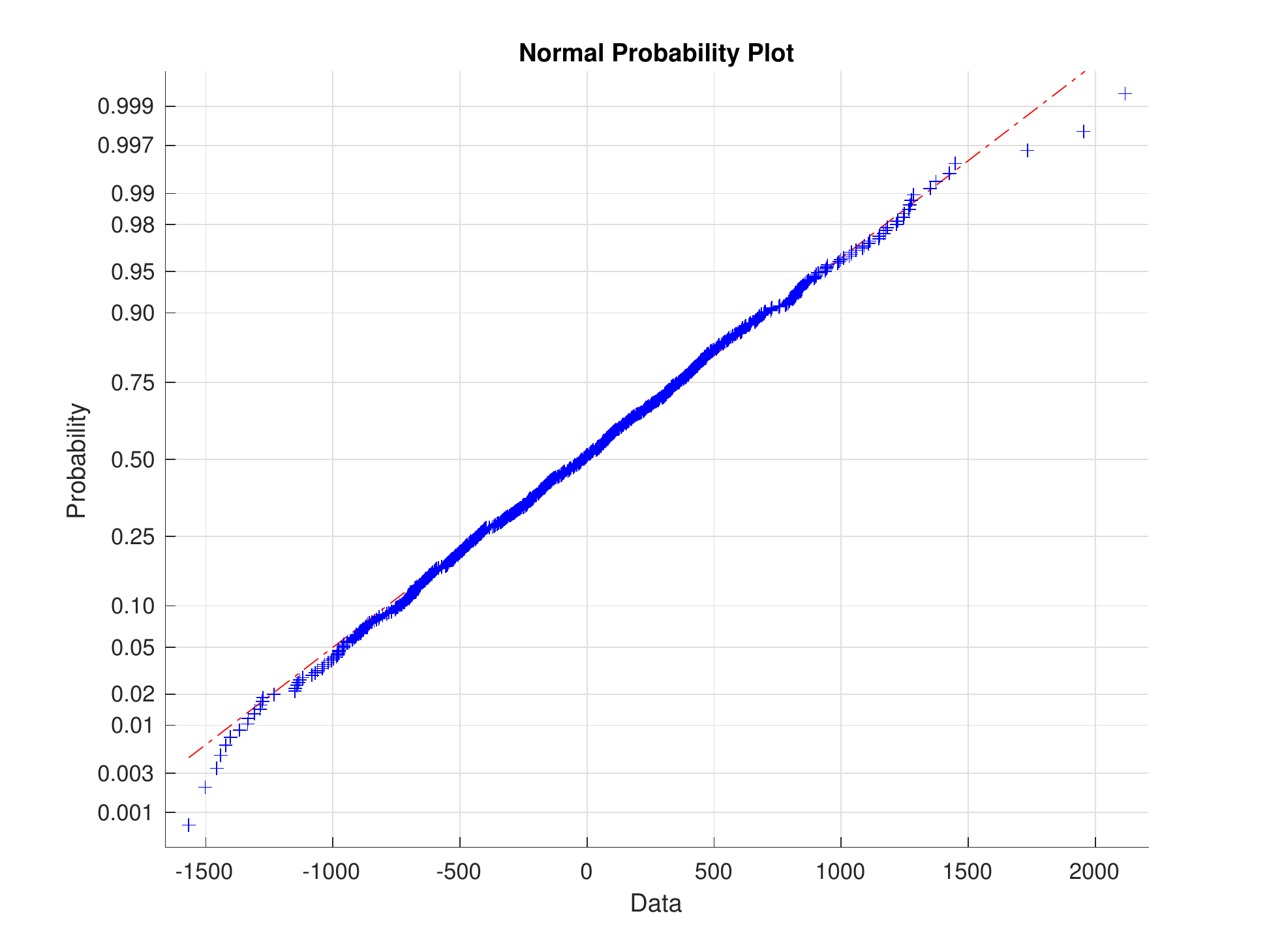}
\caption{Normal probability plot of $W_{f,N}$.}
\label{fig:Wf_norm}
\end{figure}

The terms $\mathbf{W}_f^\downarrow$ and $\mathbf{W}_f^\uparrow$ are computed on the basis of a statistical analysis of past data from the last two years of frequency deviations and assuming that the BESS under control does not influence the future frequency deviation. First, the daily profiles composed by $\mathbf{W}_f=W_{f,1},\dots,W_{f,N}$ have been calculated from hystorical data, by integrating the frequency deviations measured in a set of 24~h periods. The mean $\mu_{W,k}$ and variance $\sigma^2_{W,k}$ of such values have then been computed for all $k=1,\dots,N$. It can be observed that the set of $W_{f,k}$ values is close to normally distributed for any instant $k$. A Chi-square goodness-of-fit test on the dataset does in fact not reject the null hypothesis at the 5\% significance level. In Fig. \ref{fig:Wf_norm}, it is shown that the normal probability plot of the values assumed $W_{f,k}$ for $k=N$ (i.e. at the end of the 24 hours). We then define $W^\uparrow_{f,k}$ and $W^\downarrow_{f,k}$ for all $k$ as a function of the mean value $\mu_{W,k}$ and the standard deviation $\sigma_{W,k}$ as
\begin{align}
\begin{split}
& W^\uparrow_{f,k} = \mu_{W,k} + 1.96 \sigma_{W,k} \\
& W^\downarrow_{f,k} = \mu_{W,k} - 1.96 \sigma_{W,k}, \text{ for } k=1,\dots,N 
\end{split}
\end{align}
to have a 95\% confidence level that the realization of $W_{f,k}$ will lie between $W^\uparrow_{f,k}$ and $W^\downarrow_{f,k}$. Similarly, we can define $W^\uparrow_{f,k}$ and $W^\downarrow_{f,k}$ for any other confidence level.

\bibliographystyle{IEEEtran}
\bibliography{biblio}

\begin{IEEEbiographynophoto}
	%[{\includegraphics[width=1in,height=1.25in,clip,keepaspectratio]{figures/emilnamor}}]
	{Emil Namor} received the M.Sc. in electrical engineering from the University of Padova, Italy, and the M.Sc. in engineering from the Ecole Centrale de Lille in 2014. Since 2015, he is enrolled as a Ph.D. student at the Distributed Electrical Systems Laboratory at EPFL, Switzerland. His main research interest are modeling and control of battery storage systems.
\end{IEEEbiographynophoto}

\begin{IEEEbiographynophoto}
	%[{\includegraphics[width=1in,height=1.25in,clip,keepaspectratio]{figures/fabriziosossan}}]
	{Fabrizio Sossan}
	is an Italian citizen and was born in Genova in 1985. He got his M.Sc. in Computer Engineering from the University of Genova in 2010, and, in 2014, the Ph.D. in Electrical Engineering from DTU, Denmark. Since 2015, he is a postdoctoral fellow at the Distributed Electrical Systems Laboratory at EPFL, Switzerland. In 2017, he has been a visiting scholar at NREL, Colorado, US. His main research interest are modeling and optimization applied to power system.
\end{IEEEbiographynophoto}

\begin{IEEEbiographynophoto}
	%[{\includegraphics[width=1in,height=1.25in,clip,keepaspectratio]{figures/rachidcherkaoui2}}]
	{Rachid Cherkaoui}
	(M’05–SM’07) received the M.Sc. and Ph.D. degrees in electrical engineering from the École Polytechnique Fédérale de Lausanne (EPFL), Lausanne, Switzerland, in 1983 and 1992, respectively. He is currently a Senior Scientist with EPFL, leading the Power Systems Group. He has authored or co-authored over 100 scientific publications. His current research interests include electricity market deregulation, distributed generation and storage, and power system vulnerability mitigation. Dr. Cherkaoui was a member of CIGRE TF and WG. He is a member of technical program committees of various conferences. He was the IEEE Swiss Chapter Officer from 2005 to 2011.
\end{IEEEbiographynophoto}

\begin{IEEEbiographynophoto}
	%[{\includegraphics[width=1in,height=1.25in,clip,keepaspectratio]{figures/mariopaolone.jpg}}]
	{Mario Paolone}
	(M07-SM10) received the M.Sc. (with Hons.) and Ph.D. degrees in electrical engineering from the University of Bologna, Bologna, Italy, in 1998 and 2002, respectively. In 2005, he was appointed as an Assistant Professor in power systems with the University of Bologna, where he was with the Power Systems Laboratory until 2011. In 2010, he received the Associate Professor eligibility from the Polytechnic of Milan, Italy. Since 2011, he joined the Swiss Federal Institute of Technology, Lausanne, Switzerland, where he is currently Full Professor, Chair of the Distributed Electrical Systems Laboratory, and Head of the Swiss Competence Center for Energy Research Future Swiss Electrical infrastructure. He has authored or co-authored over 230 scientific papers published in reviewed journals and international conferences. His current research interests include power systems with particular reference to real-time monitoring and operation, power system protections, power systems dynamics, and power system transients.
	Dr. Paolone was the Co-Chairperson of the Technical Program Committees of the 9th edition of the International Conference of Power Systems Transients (2009) and the 2016 Power Systems Computation Conference. He is the Chair of the Technical Program Committee of the 2018 Power Systems Computation Conference. In 2013, he was a recipient of the IEEE EMC Society Technical Achievement Award. He has co-authored several papers that received the following awards: the Best IEEE Trans. on Electromagnetic Compatibility Paper Award in 2017, the Best Paper Award at the 13th International Conference on Probabilistic Methods Applied to Power Systems, Durham, U.K., in 2014, the Basil Papadias Best Paper Award at the 2013 IEEE PowerTech, Grenoble, France, and the Best Paper Award at the International Universities Power Engineering Conference in 2008. He is the Editor-in-Chief of the journal Sustainable Energy, Grids and Networks (Elsevier) and an Associate Editor of the IEEE Trans. on Industrial Informatics.
\end{IEEEbiographynophoto}

\end{document}